\documentstyle[11pt, aas2pp4]{article}
\tighten

\begin{document}

\def\j{\hbox{$J$}}		
\def\h{\hbox{$H$}}		
\def\k{\hbox{$K$}}		
\def\v{\hbox{$V$}}		
\def\r{\hbox{$R$}}		
\def\b{\hbox{$B$}}		
\def\bh{\hbox{$B\!-\!H$}}       
\def\vh{\hbox{$V\!-\!H$}}       
\def\vk{\hbox{$V\!-\!K$}}       
\def\bk{\hbox{$B\!-\!K$}}       
\def\hk{\hbox{$H\!-\!K$}}       
\def\Mv{\hbox{$M_V$}}               
\def\Mh{\hbox{$M_H$}}               
\def\Mk{\hbox{$M_K$}}               
\def\Mb{\hbox{$M_B$}}               

\def\micron{\hbox{$\mu m$}}                     
\def\perpix{\hbox{$\rm pix^{-1}$}}              
\def\arcsec{\hbox{$^{\prime\prime}$}}           
\def\msa{\hbox{$\rm mag~arcsec^{-2}$}}          
\def\kms{\hbox{$\rm km~s^{-1}$}}                
\def\Hfifty{\hbox{$H_0=50\rm~km~s^{-1}~Mpc^{-1}$}}   
\def\Heighty{\hbox{$H_0=80\rm~km~s^{-1}~Mpc^{-1}$}}  
\def\q00{\hbox{$q_0=0$}}                          
\def\lstar{\hbox{$L^*$}}                          
\def\rquarter{\hbox{$r^{1/4}$}}                   
\def\lesssim{\mathrel{\hbox{\rlap{\hbox{\lower4pt\hbox{$\sim$}}}\hbox{$<$}}}}
\def\gtrsim{\mathrel{\hbox{\rlap{\hbox{\lower4pt\hbox{$\sim$}}}\hbox{$>$}}}}
\def\Msun{\hbox{$M_\odot$}} 
\def\Lsun{\hbox{$L_\odot$}} 
\def\etal{et al.}         

\title{Obscuration in the Host Galaxies of Soft X-ray Selected Seyferts}
\author{Robert Simcoe\altaffilmark{1}, K. K. McLeod, Jonathan
Schachter, Martin Elvis}
\affil{Smithsonian Astrophysical Observatory, 60 Garden St.,
Cambridge, MA 02138\\
rasimcoe@phoenix.princeton.edu\\
kmcleod, jschachter, \& melvis @cfa.harvard.edu}
\altaffiltext{1}{Princeton University Observatory, Peyton Hall,
	Princeton, NJ 08544}

\authoremail{rasimcoe@phoenix.princeton.edu; kmcleod,jshachter,melvis@cfa.harvard.edu}

\begin{abstract}

We define a new sample of 96 low-redshift ($z<0.1$), soft X-ray
selected Seyferts from the catalog of the {\it Einstein} Slew Survey (Elvis
\etal~1992, Plummer \etal~1994).  We probe the geometry and column
depth of obscuring material in the host-galaxy disks using galaxian
axial ratios determined mainly from the Digitized Sky Survey.  
The distribution of host-galaxy axial ratios clearly shows a bias
against edge-on spirals, confirming the existence of a geometrically
thick layer of obscuring material in the host-galaxy planes.  Soft
X-ray selection recovers some of the edge-on objects missed in UV and
visible surveys but still results in 30\% incompleteness for Type 1's.
We speculate that thick rings of obscuring material like the ones we
infer for these Seyferts might be commonly present in early type spirals,
sitting at the Inner Lindblad Resonances of the nonaxisymmetric
potentials of the host galaxies.

\end{abstract}


\section{ Introduction}

In studies of active galactic nuclei (AGN), much attention has 
been paid to ``unified'' schemes (e.g. Antonucci \& Miller 1985;
Antonucci 1993) that explain the observed differences among the
classes of Seyferts based on our viewing angle to the nucleus.  
In these models, an optically thick torus with a
radius of about 1 pc surrounds the broad line region (BLR).  When our
line of sight intercepts this torus, the central continuum source and
the BLR are hidden from our view and we see only the Type 2 spectrum
from the narrow line region (NLR).  

Far fewer investigations have focused on obscuration on a
larger scale.  
Keel (1980) found that optically-selected (mostly Type 1)
Seyferts tend to avoid host galaxies with axial ratio $b/a<0.5$,
i.e. they are rarely found in edge-on galaxies.  By contrast, 
hard X-ray (2-10 keV) selected Seyferts show no such bias
(Lawrence \& Elvis 1982).  Lawrence \& Elvis concluded that
optically-selected samples  
must suffer biases due to  obscuration in a flattened configuration
parallel to the disk of the  host galaxy.  This and subsequent studies
indicated that the obscuring material is likely inside or at the outer
edge of the BLR (e.g. De Zotti \& Gaskell 1985), though a  more recent
look suggested that the obscuring material might extend into the
NLR (Kirhakos \& Steiner 1990a).   

To investigate this question further and to avoid the problems
intrinsic to studying heterogeneous samples, 
McLeod \& Rieke (1995) compiled statistics on several sets of AGN
selected in different wavelength regimes.  
They found an excess of face-on spirals hosting both Type 1 {\it and}
Type 2 AGN for a spectroscopically-selected sample.  Therefore, a
significant amount of dust in the galaxy plane must cover at least
part of the NLR, i.e., $\rm A_V\gtrsim$ several magnitudes out to 
$\rm r\gtrsim100~pc$.  They also found an excess of face-on spirals 
hosting soft X-ray selected AGN, and used it to estimate
the column density along the line of sight to the X-ray continuum
source, inferring $\rm N_H\sim10^{23}~cm^{-2}$ at scales $\rm r>1~pc$.  In both
regimes, the aspect ratio (i.e. height-to-thickness) of the obscuring
material is $\approx 1$. 

A recent investigation by Maiolino \& Rieke (1995) provides
preliminary evidence that this additional component of obscuration 
might explain the axial ratio
distributions and the spectral characteristics of intermediate Seyferts.
In their model, a line of sight that intercepts the plane of the galaxy but
not the inner torus would make a Seyfert 1 appear as a 1.8 or 1.9 (see
their Figure 6).  Maiolino \& Rieke (1995) also present a theoretical
model for the origin of geometrically thick, 100 pc-scale obscuring
material in the context of tidal disruption of giant molecular clouds
approaching the central compact object. 

We stress that the galaxy inclination
statistics are telling us about obscuring material that is
distinct from, and not necessarily aligned with, the canonical
torus of unification models.   
Radio maps and Hubble Space Telescope images indicate that the pc-scale
torus of the unified model is not preferentially aligned with the larger scale
galaxy disk (Schmitt \etal~1996; Wilson \& Tsvetanov 1994; Ulvestad \&
Wilson 1984).
Seyfert 2's are often found in face-on spiral disks---
and in the unified model, this means that the torus is in fact
perpendicular to the host-galaxy disk.

The strongest limit placed by McLeod \& Rieke (1995) on the 
column density in the galaxy's disk was based on a sample of AGN 
serendipitously discovered by the {\it Einstein} IPC 
and reported by Malkan, Margon, \& Chanan (1984).
However, the Malkan \etal~selection process might have
eliminated {\it a priori} edge-on galaxies for which the X-rays 
got through but for which emission lines were absorbed, thus
precluding their identification as AGN.
In this paper, we use a new, complete, well-defined, soft X-ray sample
of AGN from the {\it Einstein} Slew Survey (Elvis \etal~1992; Plummer \etal~1994).
We measure axial ratios from Digitized Sky Survey (DSS\footnote
{\footnotesize Northern hemisphere based on  photographic data of the
National Geographic Society -- Palomar Observatory Sky Survey
(NGS-POSS) obtained using the Oschin Telescope on Palomar Mountain.
The NGS-POSS was funded by a grant from the National Geographic
Society to the California Institute of Technology.  Southern
hemisphere based on photographic data obtained using The UK  Schmidt
Telescope. The UK Schmidt Telescope was operated by the Royal
Observatory Edinburgh, with funding from the UK Science and
Engineering Research  Council, until 1988 June, and thereafter by the
Anglo-Australian  Observatory.  Original plate material is copyright
(c) the Royal Observatory Edinburgh and the Anglo-Australian
Observatory.  All plates were processed into the present compressed
digital form with permission from the National Geographic Society,
Royal Observatory, and The AAO.  The Digitized Sky Survey was produced
at the Space Telescope Science Institute under US Government grant NAG 
W-2166.}) images
and from new CCD images.  This soft X-ray sample allows us 
to probe column densities in between those previously examined in
UV-excess and hard X-ray samples.

\section{Sample Selection}\label{sec-sample}

Our sample was chosen from the catalog of the {\it Einstein} Slew Survey in
the soft X-ray band of 0.2-3.5 keV (Elvis
\etal~1992, Plummer \etal~1994).  This survey has a flux limit of 
approximately $\rm 3\times10^{-12}erg~cm^{-2}~s^{-1}$   (Elvis
\etal~1992) and is well-suited for
defining a sample because of its extensive sky coverage.  It maps
about 50\% of the 
sky, and contains 809 sources, 147 of which are classified as AGN.
The objects were identified as AGN by correlation with pre-existing
catalogs, or by examination of new spectra (Perlman \etal~1996;
Huchra, Schachter, \& Remillard private communication).  

The most important requirement for our sample is that the
objects be nearby enough to resolve the host galaxy of the AGN in
DSS images, because undersampling can lead to errors in axial ratio
measurements.  We restricted our sample to
Slew Survey AGN with $z<0.1$, leaving us with 96 objects.  The mean
and median redshift for this final sample is $z=0.042$, at which
distance  a galaxy diameter of 25 kpc corresponds to an angle of
34\arcsec~($\rm H_0=80~km/s/Mpc$).  This is approximately the size of $\rm
D_{25}$, the diameter at the B=$\rm 25~mag/square~arcsec$ isophote,
for a typical spiral galaxy.

Although the Slew Survey objects have been mostly identified, there still
remain 129 objects without known optical counterparts.  
We examined DSS images of all of these fields to make sure
that there are not significant numbers of $z<0.1$ AGN among the
unidentified objects.  
We checked
within the 95\% confidence limit of the {\it Einstein} position
(2\arcmin~radius; Elvis \etal~1992) for galaxies with apparent sizes
$\geq15\arcsec$, or approximately $D_{25}$ for a $z=0.1$ galaxy.
There were only three galaxies satisfying these criteria, and only one of these
three is edge-on.  We conclude that our sample is at least 97\%
complete for $z<0.1$, and that there 
is an insignificant number of potentially active edge-on galaxies with
$z<0.1$ among the unidentified fields of the Slew Survey.

A more subtle concern in our sample selection was the identification of 
35 Slew Survey objects that were classified as ``normal'' (i.e. non-AGN)
galaxies.  We are testing for obscuring material in the galaxy's disk 
that can hide emission from the nuclear region.  
It is possible that X-rays from an AGN could penetrate this obscuring
material, whereas visible light would be attenuated enough to hide the
emission lines we 
use to identify an AGN.  If this were the case, we would
expect to observe many Slew Survey ``normal'' galaxies to be hidden
AGN seen at high inclination.
To test this possibility,  we determined axial ratios for 20 of the 26 
non-AGN spiral galaxies identified in the Slew Survey (the other 6 were too
small, nebulous, or disturbed for a reliable measurement).
The highly inclined members of this population are not
likely to be hidden AGN---the 4 objects with $b/a<0.45$ are:
the well-known starburst galaxies M82 and NGC253,
the very nearby galaxy NGC2903, and the Andromeda Galaxy (M31).
We conclude that there are not significant
numbers of AGN with edge-on galaxies hidden in this subsample.

The final AGN sample is listed in Table 1.
Catalog designations, axial ratios, and redshifts were extracted from
NED\footnote{
The NASA/IPAC Extragalactic Database (NED) is operated by 
the Jet Propulsion Laboratory and the California Institute of
Technology, under contract with the National Aeronautics and Space
Administration
} 
when available, and supplemented with data from
new spectra obtained as part of the Slew Survey follow-up.
We mark in Table 1 the 19 objects with $\rm z<0.1$ that were
not known as AGN before the Slew Survey.
Of the 96 AGN in our sample, only seven are known or probable Type 2's
(Table 1).
Such a small fraction of Seyfert 2's in a soft X-ray sample is not
surprising given the known high X-ray absorption columns through
the pc-scale torus ($N_H>10^{23}$; Awaki \etal~1991).

\section{Axial Ratios}\label{sec-analysis}

Visible images of all galaxies in the sample were extracted  from the
DSS using the
software package {\it GetImage.}\footnote
{\footnotesize {\it GetImage} was created by J.Doggett for the
Association of Universities for Research in Astronomy at the Space
Telescope Science Institute, copyright AURA 1993}. 
The digitized images had a resolution
of 1\farcs7 per pixel, with typical FWHM in the range of
$3\farcs6$.  Previous studies of galaxy inclination statistics (Binney
\& deVaucouleurs 1981) have measured isophotes out to $D_{25}$
so we tested our images to ensure we could measure isophotes at least
as far.  Comparison of northern hemisphere DSS images with 
CCD test images from the 48\arcsec~telescope at the
Fred Lawrence Whipple Observatory (FLWO) on Mt. Hopkins, AZ,  showed
that the northern DSS outer isophotes reach regions of surface
brightness corresponding to 
$B\approx25.3$ \msa. The southern hemisphere DSS plates go slightly deeper,
so all of the DSS images are of an appropriate depth for this work.
For four of the objects (denoted '**' in Table 1), we computed axial
ratios based on 
H-band images from McLeod \& Rieke 
(1994).  These objects are fairly distant and have very bright nuclei,
so the high-resolution deep images were used to obtain better accuracy.

\begin{figure}[hbt]
\plotfiddle{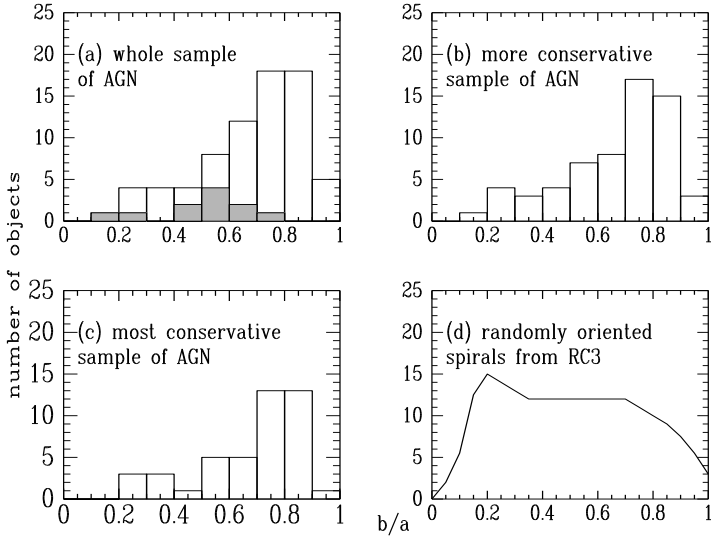}{2.5in}{0}{50}{50}{-170}{-30}
\caption[histnew.eps]{(a-c) Histograms of axial ratios for AGN 
from the {\it Einstein} Slew Survey
(samples are explained in the text).  The objects first
identified as AGN by the Slew Survey are shown as filled bars.  Even
the most conservative AGN sample shows a strong bias against edge-on
host galaxies. (d) Distribution for randomly oriented spiral galaxies
from the RC3 catalog, adapted from Figure 3 of Fasano \etal~1993 and
arbitrarily normalized for comparison.
\label{fig-hist}}
\end{figure}

The images were all processed using the {\em ellipse\/} task in  
IRAF to fit elliptical isophotes at increasing radii to each
galaxy.  The isophote fits were carried out to the level 
at which the average signal in an isophote was slightly larger than
the rms deviation of the flux in the isophote, i.e. signal-to-noise
approximately unity.  The outermost ellipses were used to
determine the axial ratio of the host galaxy.  
On average, the formal ellipticity error generated by the fitting algorithm
was $\pm0.086$, which is greater than the difference in $b/a$ between
consecutive isophotes at outer radii. The axial ratios are given in
Table 1.
For about 25\% of the sample, we were unable to measure a reliable axial
ratio.  In five cases a foreground object contaminated the galaxy
(denoted ``f'' in Table 1).  In 16 cases the object was too
small or stellar (``s''); all of these are substantially more distant than the
median redshift for the sample.  In five additional cases the object
was morphologically peculiar or interacting (``m'').  For six very 
nearby galaxies that were bigger than our images we simply adopted the
NED value of b/a.
A histogram of axial ratios of the remaining 74 galaxies is shown
in Figure \ref{fig-hist}a.  There is an obvious deficiency of
galaxies with small axial ratios.

One test of the accuracy of our fits is to compare our ratios to those
objects in the sample with previously measured axial ratios
found in NED (usually from the Third Reference Catalogue of Bright
Galaxies, deVaucouleurs \etal~1991, hereafter RC3) and listed in Table 1.  
The axial ratio agreement is excellent: the difference between our
value and NED's value has a mean of 0.02
and standard deviation of 0.12, as
expected from the ellipse fitting uncertainties.  We 
conclude that our axial ratios are accurate to $b/a\approx0.1$.  
Two of the three most discrepant points are easily reconciled:  the
multiple ``nuclei'' of interacting galaxy  1ES0655$+$542 bias the
measurement so we discard this object from our final analysis; for 
1ES0403$-$373 we noted that our resolution is poor, and hence we adopt
the smaller NED value for the analysis.  For 1ES0057$+$315, however,
we see no obvious reason for a discrepancy.  We believe our
measurement is robust and use it in our analysis.
The fraction of objects for which our axial ratios are
larger than NED's  is 
too small to change the character of the $b/a$ distribution.

\begin{figure}[hbt]
\epsscale{1.5}
\plotfiddle{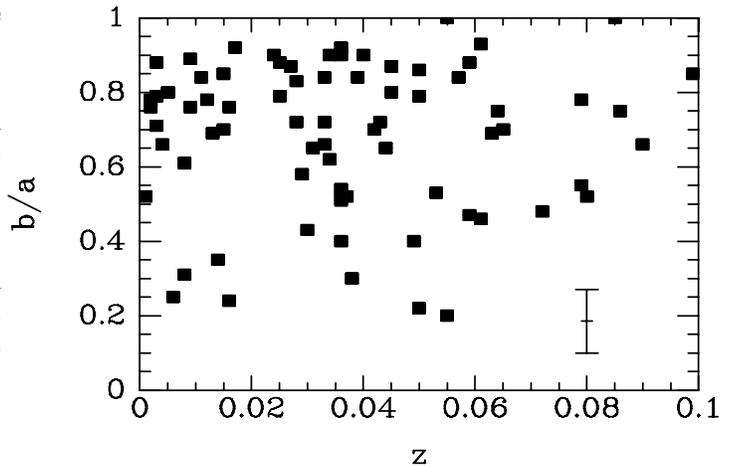}{2.25in}{0}{90}{90}{-290}{-280}
\caption[zhist.eps]{Distribution of axial ratios v. redshift for the
whole sample.  Typical $\pm 1\sigma$ error bar is shown in lower
right.  Although the bias against hosts with axial ratios 
$b/a\lesssim0.6$ appears over the entire range of z, we exclude 
objects with $z\geq0.05$ from our most conservative sample.
\label{fig-zdist}}
\end{figure}

The 1\farcs7 pixels of the DSS generally do not fully sample the 
seeing profiles of the original plates.  
To understand the effects of the POSS's resolution limits for the more
distant AGN in our sample, we 
obtained 1\farcs5 resolution CCD images at the FLWO 48'' telescope for
10 of the more distant  AGN (denoted '*' in Table 1).
The average difference in $b/a$ derived from POSS images and CCD
images was within the expected error margin of the ellipse fitting
procedure.  
However, there is still the possibility that both sets of
data undersample the galaxy images.
We do see edge-on  galaxies of similar size to the Seyfert hosts in
the DSS images, but a bright Seyfert nucleus could artificially 
inflate the axial ratio for more distant objects.
Figure \ref{fig-zdist} is a plot of the axial ratios of the galaxies in our
sample versus their redshifts.  The excess of round galaxies is seen 
for the whole range of $z$ and is not solely a distance-dependent
selection effect. However, the bias is possibly not as strong for 
$z\geq0.05$, and many of the objects in that range have already been
discarded by us because they were too compact for a reliable
measurement.  Therefore, we will define below a  ``most conservative
sample'' containing only the 44 objects with $z<0.05$.

\section{Results}\label{sec-results}

Axial ratios for the ``whole sample'' of 74 AGN for which we have measurements
are shown in Figure \ref{fig-hist}a.  For comparison, we plot in
Figure \ref{fig-hist}d the distribution derived by Fasano 
\etal~(1993) based on their
analysis of $>1000$ randomly oriented spiral galaxies drawn
from the RC3 catalog. 
The distribution shows a lack of galaxies with $b/a<0.15$, due to
the finite thickness of spiral disks.  It also shows a dearth of
galaxies with $b/a>0.8$, interpreted as an effect of intrinsically
triaxial disks.
It is immediately obvious that the bias against edge-on
hosts is strongly present
in the soft X-ray sample; the distribution
drops off considerably for $b/a \lesssim 0.5-0.6$, similar to what was
seen for UV and spectroscopic (visible) samples.  
We find from a KS test that the soft X-ray sample 
axial ratios (Table 1) are different from those for randomly oriented 
spirals---the KS probability that the two distributions are consistent
is only $2\times10^{-9}$.
There is also an apparent lack of round galaxies, i.e. galaxies with
$b/a>0.9$.

This result persists in progressively more conservative subsamples.
In Figure \ref{fig-hist}b we plot the distribution  for a ``more
conservative'' sample after removing 
12 
objects (Table 1) whose axial ratios might not be meaningful for disk galaxy
statistics because the AGN is not in a simple disk galaxy
(flagged by ``m''), suffers from foreground
contamination (flagged by ``f''), or is a strong radio source that 
could be associated with an elliptical host galaxy (flagged by ``r'').
The KS probability that the axial ratio distribution for the resulting
subsample is consistent with that for randomly oriented disks 
is $2\times10^{-7}$.
Finally we remove the 18 remaining 
objects with $z\geq0.05$ (Figure \ref{fig-hist}c). The axial ratio
distribution for this ``most conservative'' sample
of 44 objects still shows a deficit
of highly inclined galaxies, with a KS probability of 
$3\times10^{-6}$ of being consistent with that for randomly
oriented disks.

This result confirms the previous tentative result
of McLeod \& Rieke (1995) that soft X-ray selection suffers a bias
against edge-on host galaxies. 
We estimate that the soft 
X-ray sample misses $\approx 60\%$ of the AGN in galaxies with
$b/a<0.6$,
corresponding to a missing fraction of $\approx 30\%$ over all
inclinations.  These statistics apply to only the Seyfert 1's---recall
from \S\ref{sec-sample} that Seyfert 2's are not usually detected at
all in soft x-ray samples. 

The axial ratio $b/a<0.6$ corresponds to a galaxy 
inclination of $i\gtrsim55$ degrees, which determines the
opening half-angle of the obscuring material coplanar with the host
galaxy's disk and implies a distribution of material with 
a thickness approximately equal to its distance from the galaxian
center.
The missing fraction 
implies a large covering factor for the soft X-ray
absorber, $f_c\gtrsim0.6$ for $i\gtrsim55$ degrees.

The soft X-ray regime used to select the AGN used in our study
allows us to put a stringent lower limit on the column density of 
obscuring material.
For solar abundances and a typical AGN intrinsic photon
spectrum of $N(E)\propto E^{-1.7}$ (e.g. Petre \etal~1984)
a column $\rm N_H\approx10^{22}~cm^{-2}$ is
required to give optical depth $\tau=1$ over the {\it Einstein}
bandpass (Harris 1984). 
A previous hard X-ray study in the 2-10 keV band (Lawrence \& Elvis
1982), which did not detect a bias toward face-on spirals, can 
be used to set an upper limit on $\rm N_H$.  
To give $\tau=1$ for a typical $2-10$ keV X-ray detector requires a column
density of $\rm\sim2\times10^{23}~cm^{-2}$ (using the MPC response in {\it 
PIMMS}, Mukai 1993).
Hence, the obscuring material has
column density in the range $\rm 10^{22}<N_H<2\times10^{23}~cm^{-2}$.
If we assume a standard Milky Way ratio of $\rm
N_H/A_V=1.9\times10^{21}$ (Savage \& Mathis 
1979), we obtain a corresponding visual extinction $\rm A_V = 5-100$ 
magnitudes.

\section{Discussion}

The bias against finding edge-on host galaxies seen in this
soft X-ray selected AGN sample is detected at high confidence.  However,
it appears to be weaker than the biases seen in UV and visible samples.
We have quantified the differences among axial ratio distributions of 
Seyferts selected in various wavelength regimes using the KS statistic
to test the null hypothesis that two distributions are drawn from the
same parent distribution.  
Hard X-ray selection is represented by the complete, though small,
hard X-ray sample 
described in Piccinotti \etal~(1982) with axial ratios compiled by 
Kotilainen \etal~(1992).
Soft X-ray selection is represented by the Slew Survey ``most
conservative'' sample in this paper.
UV selection is represented by the Cheng (1985) sample of mostly
Markarian Seyfert 1's, with the axial ratios taken from Zitelli
\etal~(1993) and supplemented by 
us using values from the literature and DSS images.
Visible spectroscopic selection is represented by the Huchra \& Burg
(1992) CfA Seyfert sample with axial ratios from McLeod \& Rieke
(1995).  

\begin{figure}[hbt]
\plotfiddle{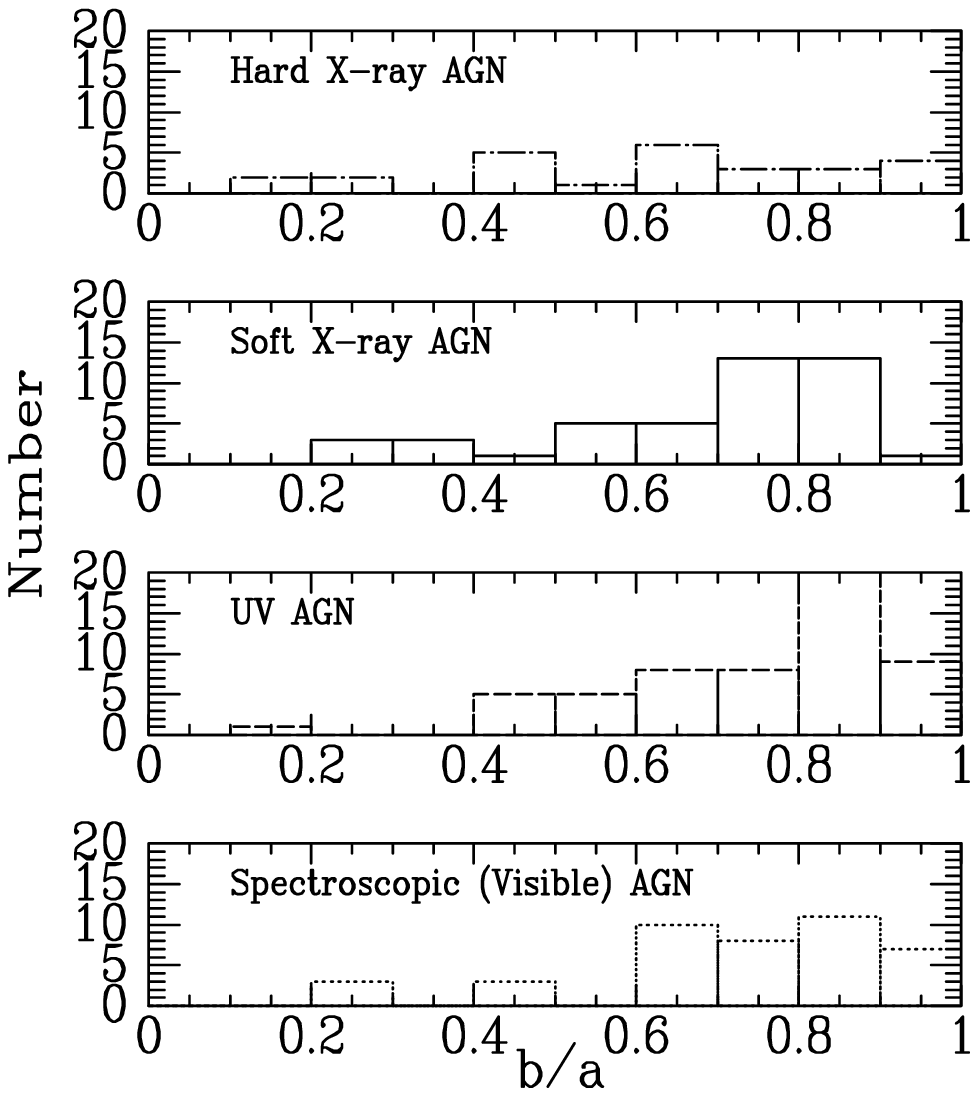}{2.25in}{0}{70}{70}{-200}{-200}
\plotfiddle{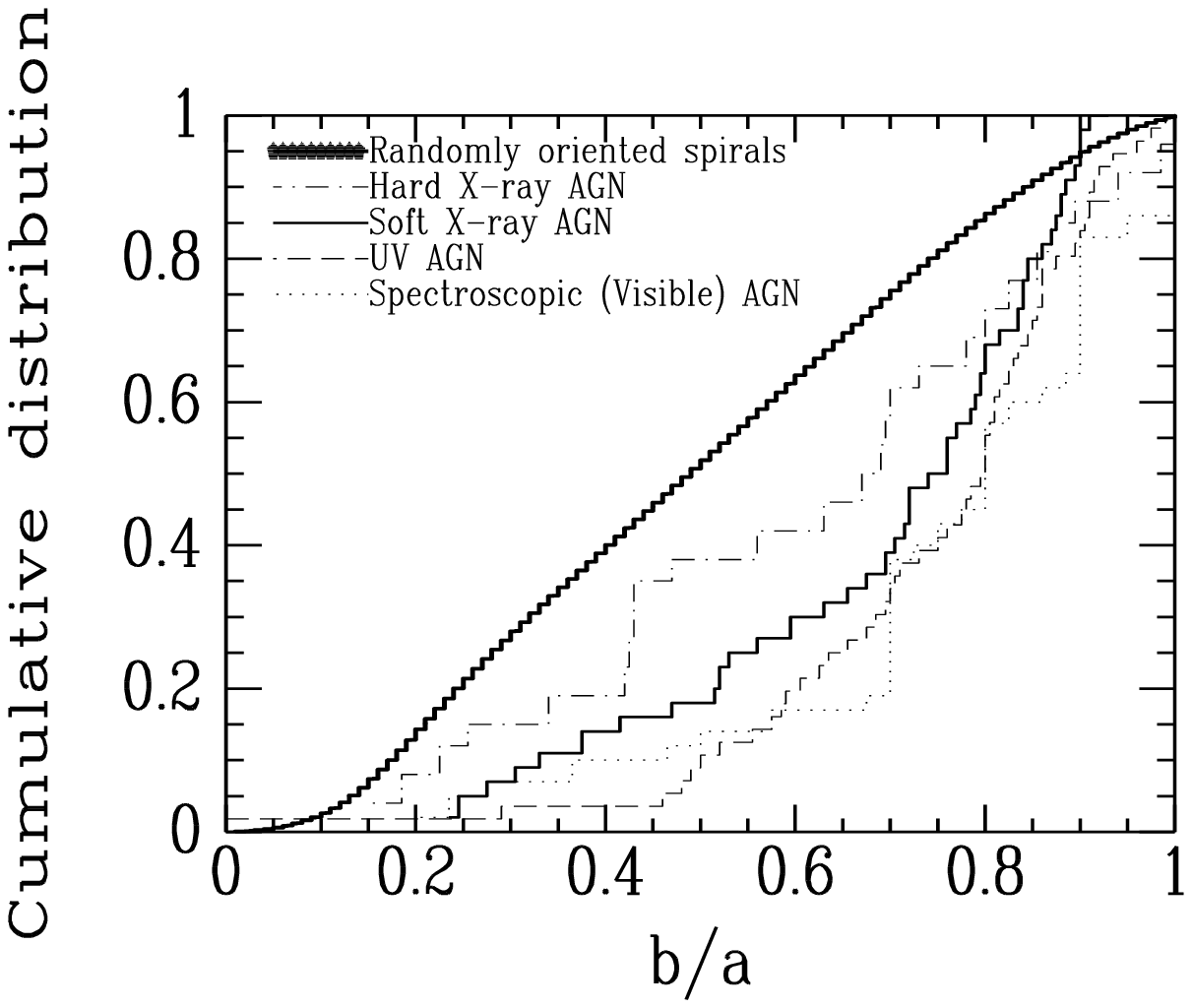}{2.25in}{0}{60}{60}{-170}{-80}
\caption[cumnew.eps]{Histograms and cumulative distributions of axial ratios for 
samples described in the text.
\label{fig-cum}}
\end{figure}

We compare these samples with each other and with the 
distribution that would be expected from a sample of randomly oriented
spirals, and we summarize the results in Table \ref{tab-ks} and 
Figure \ref{fig-cum}.  
These tests confirm that only the hard X-ray sample distribution can
be consistent with a distribution of randomly oriented disk galaxies,
and that all other samples are 
biased against edge-on galaxies with KS probability $<3\times10^{-6}$.
The KS test cannot rule out that the soft X-ray and UV samples might
be drawn from the same parent population.  However, there are two suggestions
that the soft X-ray sample is less biased than the UV sample.
First, whereas the soft X-ray distribution might be consistent with the hard
X-ray distribution (KS probability $P=0.358$), the UV
distribution is less likely to be ($P=0.043$).  A
definitive test awaits a larger hard X-ray sample.
Second, the new AGN discovered by
the Slew Survey (shaded area in Figure \ref{fig-hist}a) tend to have
axial ratios in the low $b/a$ half of the 
distribution, implying that soft X-ray selection has found a 
fraction of the AGN that are missing from UV (and visible) surveys.
It is therefore likely that the column density of the absorbing
material is close to the lower limit derived above, 
$\rm N_H\approx10^{22}~cm^{-2}$, or $\rm A_V \approx 5$ mag, at least
for inclinations near 55 degrees.

The obscuring material must lie outside the soft X-ray emitting
region of the AGN, but this is not a strong constraint.  A
substantial fraction (20-30\%) of the soft X-ray 
emission of some Seyfert galaxies is found to be extended on a kpc
scale (Morse 
\etal~1995).  However, in these galaxies the observed direct soft X-ray
continuum has been suppressed by nuclear absorption.  The extended
emission is no more than a few percent of the unabsorbed continuum
flux, which is compact on sub-pc scales.
McLeod \& Rieke (1995) have already shown that material with at least
several magnitudes of visual extinction 
and with the same orientation and thickness-to-height ratio hides at
least some of the NLR and so must lie at $\rm r\sim 100~pc$ from the
AGN's center.
It is likely that the X-ray and UV absorbing components are part of a
continuous 
distribution of gas/dust clouds.  In fact, if the column density to the 
X-ray source is close to our lower limit as suggested above, then most
of the soft X-ray 
absorption can be accounted for by the same, 100 pc-scale material that
hides the NLR.

Maiolino \& Rieke (1995) have suggested that some Seyferts of Type 1.8-1.9
are actually Seyfert 1's viewed through such a thick ``outer torus''
in the plane of the host galaxy.
Obtaining high-quality visible spectra for the AGN discovered by the
Slew Survey could help us to confirm this trend.  To investigate the 
question further, though, we will need to find more AGN in edge-on
galaxies. Hard X-ray selection and infrared selection remain good
ways to locate the missing objects.  Kirhakos \& Steiner
(1990a,b) combined hard X-ray and far-IR selection to locate new examples of
AGN in edge-on hosts.  However, if the amount of obscuring material is
close to $A_V \approx 5$ mag as we suggest, then the missing objects
should be easily visible even in the near-IR where the extinction is
lower by a factor of $\sim10$.  In addition, 
high-quality nuclear spectra in visible light can pick out
partially-hidden AGN in edge-on galaxies in the nearest galaxies,
where the sensitivity to nuclear emission is high because of their
proximity (Maiolino \& Rieke 1995; Ho 1996).

We note that a related result, namely a bias against edge-on host
galaxies,  has recently been found by Barth \etal~(1996) for UV-bright
LINERS.  They find that the visibility drops for host inclinations
$>65$ degrees, similar to that found for Seyferts.  The good spatial
resolution of the images (from WFPC2) shows dust lanes well inside 
the central kpc of the galaxies with several
magnitudes of extinction (at 2200\AA).  Similarly,
an HST UV imaging survey of an unbiased sample of 110 nearby galaxies
has turned up a tendency for UV ``null emitters'' to appear nearly
exclusively in galaxies with inclination $>60$ degrees, suggesting that the
host-galaxy disks obscure what would otherwise be seen as UV ``weak
emitters'' (Maoz \etal~1996).

As pointed out in McLeod \& Rieke (1995) and Maiolino \& Rieke (1995),
direct evidence for a 100 pc-scale, geometrically thick distribution
of molecular gas like the one we infer has been seen in several
nearby AGN (e.g. Tacconi \etal~1994; Bergman \etal~1992; Israel
\etal~1990; Rydbeck \etal~1993).
The prevalence of this bias for samples of active and even mildly
active nuclei suggests that a geometrically and optically thick
distribution of gas and dust could be a common feature of the
early-type disk galaxies in which they are generally found.
The obscuring material would be more difficult to detect in galaxies
where there is no bright, compact nucleus, especially given the
required spatial resolution: $\rm 50~pc$ corresponds to 
$\rm \theta\approx0\farcs5\times(20~Mpc)/d$.  
A high-resolution CO aperture synthesis study of the nearby,
edge-on spiral NGC891 shows evidence for a relatively thick nuclear disk
of molecular gas with $\rm r\sim225~pc$ in a normal galaxy (Scoville
\etal~1993).  The disk has thickness $\rm h\sim160~pc$, within a
factor of 1.5 as geometrically thick as we infer 
from the AGN axial ratios, and contains enough material to satisfy
the optical depth requirement.  There is so far no evidence that
this molecular disk has a central hole to provide the required opening angle.

Maiolino \& Rieke (1995) speculate that, for Seyferts, a distribution
of obscuring material with a central hole might result from the tidal
disruption of molecular clouds as they migrate inwards from the
large-scale galaxy disk into the potential of the compact central object.  
Here we suggest a complementary model that naturally accounts for a
central hole, namely that the gas is part of a nuclear ring of the
type often associated with bars or oval distortions in early-type
spirals.  Nuclear star-forming rings have sizes on the order of
several hundred pc and are probably associated with
the Inner Lindblad Resonances (ILRs) of the potential (Buta \&
Crocker 1993; Kenney 1996).  Numerical simulations of 
barred galaxies show that the buildup of central mass results in the 
formation of a strong ILR.  The gas transported inwards via radial
orbits in the large-scale bar or oval distortion collects at the ILR,
and even the cold gas there can have a geometrically thick
distribution because of vertical resonances (Friedli \& Benz 1993).

If this interpretation is correct, then we would expect the large
scale galaxy disks to show evidence of the nonaxisymmetric potential.
Indeed, the axial ratio distribution for the Slew Survey sample
shows a strong deficit of galaxies with $b/a>0.9$ indicating a lack
of round galaxies (Figure \ref{fig-hist}).  Similar deficits
were noted by Malkan, Margon, \& Chanan (1984) for their soft X-ray
sample, and by Maiolino \& Rieke for their spectroscopic sample. The UV-excess
sample in Figure \ref{fig-cum} shows a similar but weaker bias. 
Because $b/a$ cannot exceed unity, the effect of measurement
uncertainties will tend to create a deficit in the last bin, but this
effect should be relatively small given the accuracy of our measurements.
However, the distributions for normal spiral galaxies also show a bias against
perfectly round systems (Figure \ref{fig-hist}d; Binney \&
deVaucouleurs 1981; Fasano \etal~1993), and it is likely that disks of
non-active galaxies are triaxial (Binney \& deVaucouleurs 1981; Fasano
\etal~1993; Rix \& Zaritsky 1995).  We cannot determine 
from our data whether the Seyfert hosts are less often perfectly round
than those of nonactive spirals, but the observed distribution
suggests that this might be an interesting topic for future study. 

\section{Conclusions}

We have defined a complete sample of {\it Einstein} Slew Survey AGN
with $z<0.1$.   This sample comprises  96 mostly Type 1 AGN
including 19 objects at $z\approx0.05$ not previously
known to be active.  The distribution of host-galaxy axial 
ratios clearly shows a bias toward face-on spirals, confirming the 
existence of a geometrically thick layer of obscuring material in the
host-galaxy planes.  Combining this and other studies, we infer that 
this material is likely to be $\sim100$ pc from the nucleus with a thickness
of $\sim100$ pc and column density of $\rm
N_H=(1-20)\times10^{22}~cm^{-2}$, probably nearer the low end of this
range for lines of sight near the edge of the $\sim55-60$ degree
opening half-angle. 
The covering fraction of the soft X-ray absorbing material within this
thick distribution is large, about 60\%.  Soft X-ray selection
appears to recover some of the edge-on objects missed in  UV and
visible surveys 
but still results in an overall 30\% incompleteness for Type 1's.  We
speculate that thick rings of obscuring material like the ones we infer for
Seyferts might be commonly present in early type spirals, sitting at the
Inner Lindblad Resonances of the nonaxisymmetric potentials of the host
galaxies.

\acknowledgements

RAS thanks Smita Mathur for helpful discussions.
KKM gratefully acknowledges financial support from 
NASA Grant NAGW-3134 (B. Wilkes, PI) and thanks Luis Ho for useful
discussions.  This work was also supported in part 
by NASA Grant NAG5-3066(ADP) and the SAO Summer Intern Program.

\begin{deluxetable}{clllllc}
\footnotesize
\tablecaption{Einstein Slew Survey AGN with $z < 0.1$}

\tablehead{
 {1ES Number}
&{$b/a$}
&{Reason}
&{$b/a$}
&{$z$}
&{Other Designations}
&{Seyfert Type\tablenotemark{2}}
\\ &{This Paper}
&{Excluded\tablenotemark{1}}
&{NED}
&{NED\tablenotemark{2}}
&{NED}
    }

\startdata
               
1ES0003+199&0.88&\nodata&1.0&0.025&Mkn 335&1.0  \nl
1ES0008+107&\nodata&sr&\nodata&0.089&III Zw 2, Mkn 1501&1.0 \nl
1ES0048+291&0.90&\nodata&1.0&0.036&UGC 524&1.0  \nl
1ES0050+124&0.93&\nodata&1.0&0.061&1 Zw 1, Mkn 1502&NLSy1 \nl
1ES0057+315&0.85&\nodata&0.5&0.015&Mkn 352&1.0  \nl
1ES0121$-$590&0.87&\nodata&0.87&0.045&FAIR 9&1.0  \nl
1ES0124+189&0.92&\nodata&0.83&0.017&Mkn 359&NLSy1 \nl
1ES0138+391&0.52&r&\nodata&0.080&B2 0138+39B&1.0 \nl
1ES0152+022\tablenotemark{3}&\nodata&s&\nodata&0.08\tablenotemark{\dag}&\nodata&1.0\nl
1ES0206+522&\nodata&s&\nodata&0.049&GPX 02&1.0 \nl
1ES0212$-$010&0.87&\nodata&0.9&0.027&NGC 863, Mkn 590&1.5  \nl
1ES0225+310&0.24&\nodata&0.21&0.016&NGC 931, Mkn 1040&1.5  \nl
1ES0232$-$090&\nodata&m (pec)&\nodata&0.043&NGC 985&1.5 \nl
1ES0235+016&0.90&\nodata&0.9&0.024&NGC 1019&1.0  \nl
1ES0238+069&0.72&\nodata&\nodata&0.028&Mkn 595&1.0  \nl
1ES0240$-$002&0.88&\nodata&0.84&0.003&NGC 1068,M 77&2.0  \nl
1ES0241+622&\nodata&sr&\nodata&0.044&4U 0241+61&1.0 (QSO) \nl
1ES0328+054\tablenotemark{3}&\nodata&s&\nodata&0.044\tablenotemark{\dag}&\nodata&NLSy1\nl
1ES0339$-$214&\nodata&f&0.7&0.015&MS, ESO548-G0810&\nodata \nl
1ES0403$-$373\tablenotemark{3}&0.49\tablenotemark{5}&\nodata&0.2&0.055&FAIR 1119&1.5 \nl
1ES0414+379&\nodata&sr&\nodata&0.048&3C 111.0&1.0 \nl
1ES0418$-$550&\nodata&\nodata&0.80&0.005&NGC 1566&1.5  \nl
1ES0429$-$537&0.90&m (Epec)&\nodata&0.040&FAIR 303&1.0  \nl
1ES0430+052&0.66&r&0.75&0.033&Mkn 1506, 3C120&1.0 \nl
1ES0435$-$472\tablenotemark{3}&0.79&\nodata&\nodata&0.05\tablenotemark{\dag}&\nodata&1.2 \nl
1ES0439$-$085\tablenotemark{3}&\nodata&s&\nodata&0.05\tablenotemark{\dag}&\nodata&1.2-1.5 \nl
1ES0459+034&0.76&\nodata&\nodata&0.016&GHIGO, MS 0459.5+0327&1.5  \nl
1ES0513$-$002&0.84&\nodata&0.72&0.033&Mkn 1095, AKN 120&1.0  \nl
1ES0518$-$458&0.62&r&0.75&0.034&Pic. A&1.0 \nl
1ES0545$-$336\tablenotemark{3}&0.43&\nodata&\nodata&0.03\tablenotemark{\dag}&\nodata&1.8 \nl
1ES0639$-$756\tablenotemark{3}&0.66&\nodata&\nodata&0.09\tablenotemark{\dag}&\nodata&1.8 \nl
1ES0655+542&0.65&m (interacting)&0.42&0.044&Mkn 374&1.5  \nl
1ES0702+646&0.78&\nodata&\nodata&0.079&VII Zw 118&1.0  \nl
1ES0752+393&0.90&\nodata&0.88&0.034&Mkn 382&1.0  \nl
1ES0804+761&0.85&\nodata&\nodata&0.099&PG0804+761&1.0 (QSO) \nl
1ES0818+544&\nodata&s&\nodata&0.086&MS, IRAS F08187+5428&\nodata \nl
1ES0844+349&0.75\tablenotemark{**}&\nodata&\nodata&0.064&PG0844+349&1.0 (QSO)\nl
1ES0849+080&0.69\tablenotemark{**}&\nodata&\nodata&0.063&MS0849.5+0805&1.0  \nl
1ES0915$-$118&\nodata&fr&1.0&0.055&Hydra A&RG \nl
1ES0915+165&0.58&\nodata&0.57&0.029&Mkn 704&1.5  \nl         
1ES0921+525&\nodata&m (interacting)&0.4&0.036&Mkn 110&1.0  \nl
1ES0923+129&0.83&\nodata&0.85&0.028&Mkn 705&1.0  \nl
\tablebreak
1ES0942+098&0.69&\nodata&\nodata&0.013&MS 0942.8+0950&1.9  \nl
1ES0943$-$140&\nodata&\nodata&0.31&0.008&NGC 2992&2.0  \nl
1ES0951+693&\nodata&\nodata&0.52&0.001&M 81, NGC 3031&1.9  \nl
1ES1020+201&\nodata&\nodata&0.66&0.004&NGC 3227&1.5  \nl
1ES1022+519&0.80&\nodata&\nodata&0.045&Mkn 142&1.0  \nl
1ES1103+728&0.76&\nodata&0.76&0.009&NGC 3516&1.2  \nl
1ES1136$-$374&0.89&\nodata&0.89&0.009&NGC 3783&1.2  \nl
1ES1136+342&0.72&\nodata&0.83&0.033&KUG 1136+342&1.5 \nl
1ES1141+799\tablenotemark{3}&0.70&\nodata&0.63&0.065\tablenotemark{\dag}&UGC 6728&1.2 \nl 
1ES1149$-$110&0.40&\nodata&\nodata&0.049&PG1149$-$110&1.0  \nl
1ES1155+557&0.79&\nodata&0.81&0.003&NGC 3998&1.0/LINER?   \nl
1ES1200+448&0.76&\nodata&0.75&0.002&NGC 4051&1.5  \nl
1ES1208+396&\nodata&\nodata&0.71&0.003&NGC 4151&1.5  \nl
1ES1211+143&1.0\tablenotemark{**}&\nodata&\nodata&0.085&PG1211+143&1.0 (QSO/NLSy1) \nl
1ES1215+300&0.78&\nodata&0.8&0.012&NGC 4253, Mkn 766&NLSy1 \nl
1ES1219+755&\nodata&f&\nodata&0.070&Mkn 205&1.0 \nl
1ES1235+120&0.80&\nodata&0.79&0.005&NGC 4579&1.9  \nl
1ES1238$-$332\tablenotemark{3}&0.22&\nodata&0.2&0.050\tablenotemark{\dag}&ESO 381-7&1.5  \nl
1ES1244+026&\nodata&s&\nodata&0.048&PG1244+026&1.0 \nl
1ES1249$-$131&\nodata&m (interacting)&\nodata&0.014&IRAS 1249-13&1.5 \nl
1ES1257+286&\nodata&s&\nodata&0.092&X COM&1.0  \nl
1ES1320+084&0.86&\nodata&\nodata&0.050\tablenotemark{\dag}&Mkn 1347&1.0  \nl
1ES1322$-$427&\nodata&mr (S0pec)&0.78&0.002&Cen A&2.0 \nl
1ES1323+717\tablenotemark{3,4}&0.48&\nodata&\nodata&0.072&Z 1323+7145&2.0? \nl
1ES1324$-$268\tablenotemark{3}&\nodata&m (Irr)&\nodata&0.050\tablenotemark{\dag}&ESO 509-14&\nodata \nl
1ES1333$-$340&0.61&\nodata&0.6&0.008&MCG-6-30-15&1.0  \nl
1ES1346$-$300&0.35&\nodata&0.28&0.014&IC 4329A&1.0  \nl
1ES1351+400&\nodata&s&\nodata&0.062&MS 1351.6+4005&\nodata \nl
1ES1351+695&0.65&\nodata&0.56&0.031&Mkn 279&1.0  \nl
1ES1410$-$029&0.25&\nodata&0.32&0.006&NGC 5506, Mkn 1376&1.9  \nl
1ES1426+015&0.75\tablenotemark{**}&\nodata&\nodata&0.086&Mkn 1383, PG 1426+015&1.0 (QSO)\nl
1ES1501+106&0.92\tablenotemark{~*}&m (E)&\nodata&0.036&Mkn 841&1.5  \nl
1ES1518+593&0.55\tablenotemark{~*}&\nodata&\nodata&0.079&SBS 1518+593&1.0  \nl
1ES1539+187\tablenotemark{3}&\nodata&s&\nodata&0.070\tablenotemark{\dag}&\nodata&1.2-1.5 \nl
1ES1615+061&0.30\tablenotemark{~*}&\nodata&\nodata&0.038&IRAS 16154+0611&1.0\nl
1ES1618+411\tablenotemark{3}&0.54\tablenotemark{~*}&\nodata&0.75&0.036\tablenotemark{\dag}&KUG 1618+410&1.2 \nl
1ES1622+261&\nodata\tablenotemark{*}&m (interacting)&\nodata&0.040&EXO, IRAS F16221+2611&2.0? \nl
1ES1659+341\tablenotemark{3}&\nodata&s&\nodata&0.09\tablenotemark{\dag}&\nodata&1.5 \nl
1ES1702+457&0.47&\nodata&\nodata&0.059\tablenotemark{\dag}&IRAS 17020+4544&NLSy1  \nl
1ES1720+309&0.72\tablenotemark{~*}&\nodata&0.63&0.043&Mkn 506&1.5  \nl
1ES1739+518&0.46\tablenotemark{~*}&\nodata&\nodata&0.061&E1739+518&1.0  \nl
1ES1743+480\tablenotemark{3}&0.53&\nodata&\nodata&0.053\tablenotemark{\dag}&IRAS F17437+4803&NLSy1\nl
1ES1817+537\tablenotemark{3}&\nodata\tablenotemark{*}&s&\nodata&0.08\tablenotemark{\dag}&IRAS 18171+5342&1.0 \nl
\tablebreak
1ES1833+326&0.88\tablenotemark{~*}&r&\nodata&0.059&3C 282.0&1.0 \nl
1ES1845+797&0.84&r&\nodata&0.057&3C 390.3&1.0 \nl
1ES1927+654\tablenotemark{3}&\nodata&f&\nodata&0.017\tablenotemark{\dag}&\nodata&1.9-2 \nl
1ES1934$-$063&0.84&\nodata&\nodata&0.011&SS 442&2.0? \nl
1ES1957+405&\nodata\tablenotemark{*}&fr&\nodata&0.058&Cyg A&2.0 \nl
1ES2055+298\tablenotemark{3}&0.51&\nodata&\nodata&0.036\tablenotemark{\dag}&\nodata&NLSy1 \nl 
1ES2137+241\tablenotemark{3}&0.52&\nodata&\nodata&0.037\tablenotemark{\dag}&\nodata&NLSy1? \nl
1ES2240+294&0.79&\nodata&0.57&0.025&ARK 564&NLSy1  \nl
1ES2251$-$178&\nodata&s&\nodata&0.068&MR 2251-178&1.0(QSO) \nl
1ES2254$-$371&0.84&\nodata&\nodata&0.039&MS 2254.9-3712&\nodata \nl
1ES2304+042&0.70&\nodata&\nodata&0.042&PG2304+042&1.0  \nl

\tablenotetext{1}{Object excluded from one or more subsamples for 
the following reason: 
(f) foreground contamination;
(m) morphological type not spiral (type information in parentheses);
(r) radio source;  
(s) too small for reliable determination of $b/a$. 
See text.}
\tablenotetext{2}{From NED and other literature unless marked with 
a \dag, in which case derived from Slew Survey spectra.  Note that the
spectra from which the Seyfert types were identified are not always of
sufficient quality to distinguish among subtypes.}
\tablenotetext{3}{Not known as a Seyfert before the Slew Survey.}
\tablenotetext{4}{1ES1323+717 was located outside the 2\arcmin~95\% error
	circle of the Slew Survey.}
\tablenotetext{5}{Our axial ratio measurment for 1ES0403-373 is likely an
upper limit; we adopt the NED value for statistical use.}
\tablenotetext{*}{CCD image obtained for this object.}
\tablenotetext{**}{IR image used to measure axial ratio.}

\enddata

\end{deluxetable}

\begin{deluxetable}{lllll}
\footnotesize
\tablecaption{KS Probabilities Among AGN Samples\label{tab-ks}}
\tablewidth{0pt}
\tablehead{
\colhead{~} & \colhead{Hard X-ray} & \colhead{Soft X-ray} & 
\colhead{UV}  & \colhead{Visible} \nl
\colhead{~} & \colhead{26 objects} & \colhead{44 objects} &\colhead {56 objects} &\colhead{42 objects} 
}
\startdata

Random orientations &0.045   &$3\times10^{-6}$ &$7\times10^{-11}$ & $2\times10^{-12}$ \nl 
Hard X-ray          &\nodata &0.358         & 0.043    & 0.011      \nl
Soft X-ray          &\nodata &\nodata       & 0.669    & 0.054      \nl
UV                  &\nodata &\nodata       & \nodata  & 0.349      \nl

\enddata
\end{deluxetable}

\end{document}